\documentclass[sigconf]{acmart}

\copyrightyear{2024}
\acmYear{2024}
\setcopyright{rightsretained}
\acmConference[ASE '24]{39th IEEE/ACM International Conference on Automated Software Engineering }{October 27-November 1, 2024}{Sacramento, CA, USA}
\acmBooktitle{39th IEEE/ACM International Conference on Automated Software Engineering (ASE '24), October 27-November 1, 2024, Sacramento, CA, USA}
\acmDOI{10.1145/3691620.3695304}
\acmISBN{979-8-4007-1248-7/24/10}

\setcitestyle{nosort}

\begin{document}

\title{Build Issue Resolution from the Perspective of Non-Contributors}

\author{Sunzhou Huang}
\email{sunzhou.huang@utsa.edu}
\affiliation{%
  \institution{University of Texas at San Antonio}
  \streetaddress{One UTSA Circle}
  \city{San Antonio}
  \state{Texas}
  \country{USA}
  \postcode{78249}
}
\author{Xiaoyin Wang}
\email{xiaoyin.wang@utsa.edu}
\affiliation{%
  \institution{University of Texas at San Antonio}
  \streetaddress{One UTSA Circle}
  \city{San Antonio}
  \state{Texas}
  \country{USA}
  \postcode{78249}
}
\begin{abstract}
Open-source software (OSS) often needs to be built by roles who are not contributors. Despite the prevalence of build issues experienced by non-contributors, there is a lack of studies on this topic. This paper presents a study aimed at understanding the symptoms and causes of build issues experienced by non-contributors. The findings highlight certain build issues that are challenging to resolve and underscore the importance of understanding non-contributors' behavior. This work lays the foundation for further research aimed at enhancing the non-contributors' experience in dealing with build issues.
\end{abstract}


\maketitle

\section{Introduction}
Open-source software (OSS) frequently requires building from source code. This ``build'' process often encompasses several steps, including compiling the source code, resolving dependencies, packaging, testing, performing static analysis, generating documentation, and preparing for deployment. The primary roles of those engaged in these building activities are those of project contributors, who manage repositories and commit code. However, there are also numerous non-contributors who need to build the project for their own goals. Examples of such non-contributors are as follows:

\textbf{Users:} Not all OSS projects provide pre-built installation packages. Even when available, these packages may not be compatible with a user's operating system (OS) or local environment. So some users may need to build projects from source code in their local environments. Advanced users might also prefer non-default configurations, utilize new features not yet included in a stable release, or compile directly from the source code for security reasons, such as in blockchain scenarios~\cite{Bitcoin}.

\textbf{Learners:} Individuals who are new to OSS development and are in the learning phase also engage in building OSS. They may start by compiling and experimenting with the source code to understand the project structure and functionality better. This hands-on experience is essential for their growth as OSS contributors.

\textbf{Potential contributors:} Individuals who wish to propose feature improvements or bug fixes for OSS must also build the project initially. This step is necessary before they can validate their revisions and submit a pull request. However, because OSS often has diverse build environments, these potential contributors frequently lack the required local build setups. Setting up these environments themselves can pose its own set of challenges.

In this paper, within the context of the OSS build process, we use the term ``non-contributor'' to refer to roles that lack familiarity with the build environment of the target OSS and need to set up their local environment. These roles may have varying degrees of background knowledge about the target OSS, but they all share the common objective of building the OSS from its source code.

Since non-contributors are often unfamiliar with the build environments of the OSS projects they aim to work on, many encounter challenges in this process. On Stack Overflow (SO), filtering by the keywords ``not build'' yielded 6,760 out of 43,140 (15.7\%), 7,374 out of 95,275 (7.73\%), and 10,336 out of 73,559 (14.1\%) relevant SO questions for three well-known open-source projects: Tomcat, Hibernate, and OpenCV, respectively~\cite{SO}. Since OSS contributors are unlikely to address internal build issues on Stack Overflow (preferring internal issue tracking systems like GitHub Issues or JIRA), it is reasonable to infer that most of these SO questions are from non-contributors experiencing build issues in their local environments. These numbers highlight the prevalence and difficulty of build issues faced by non-contributors. Another illustrative example is evident in computer science (CS) classes, where students frequently encounter initial project hurdles. These difficulties typically arise from their inability to build or install the frameworks or tools on their own systems.

Despite the prevalence of build issues from non-contributors, existing research has largely overlooked these challenges, focusing primarily on the perspective of contributors, such as the developers of the project being built. For instance, extensive studies have been conducted on the effort required to maintain build scripts~\cite{mcintosh2011empirical,mcintosh2015large}, the categorization and distribution of bugs in build scripts~\cite{barrak2021builds,shridhar2014qualitative,xia2014empirical}, the patterns of fixing build scripts~\cite{lou2020understanding,zhao2014empirical}, and how build failures occur in new integration scenarios such as Continuous Integration (CI) chains~\cite{zolfagharinia2017not} or Docker environments~\cite{wu2020empirical}. It is important to note that build issues experienced by non-contributors may have different root causes compared to those experienced by contributors. The latter are often caused by flaws in the build scripts or code of the project being built, while the former are likely related to the local environments of the non-contributors, assuming that most OSS projects are released with validated build scripts. One potential reason why build issues experienced by non-contributors are not studied may be due to data collection challenges. Most existing studies gather data from version histories and build logs, while non-contributors' experienced build issues are never systematically tracked or recorded.

In this paper, we present a study to investigate build issues experienced by non-contributors, aiming to answer the prevalent question, \textbf{Why does the OSS project not build on my machine?}. Our study tracks the behaviors of senior-year and graduate computer science students as they undertake build tasks for various OSS projects. We believe these students are representative subjects for our study, as they currently occupy learner roles for the target OSS projects and could potentially transition to user or potential contributor roles with appropriate guidance. All these students have a certain level of background in information technology (most have some internship and working experience), which positions them for a more advanced user role in utilizing the target OSS projects. We have designed these build tasks as a course project for a cross-listed software engineering course, where we instructed 31 students to build 12 different OSS projects in 6 programming languages (PLs) in order to explore the symptoms and causes of build issues experienced by non-contributors.

\vspace{-.2cm}
\section{Related Works}
\label{sec:related}

The research efforts most related to our research are user studies on software build tasks. Kwan et al.~\cite{kwan2011does} studied developers from IBM to find out whether team composition and coordination may have an impact on software build success. Dawns et al.~\cite{downs2012ambient} studied the operation of a build team to evaluate an automatic build management tool that enforces the build failure handling process. Philips et al.~\cite{phillips2014understanding} studied software building teams at Microsoft and found that most challenges are on the social aspects of the team. Kerzazi et al.~\cite{kerzazi2014automated} studied 3,214 software builds and found that 17.9\% of builds failed and more than 300 man-hours were cost to fix them. Hilton et al.~\cite{hilton2016continuous} performed interviews with 16 developers to find out their opinion on whether the CI process may enhance software productivity. Vassallo et al.~\cite{vassallo2020every} performed a study with 17 participants to find out how well developers can take advantage of BART, a tool for summarizing build failure reasons. Different from our research, all these studies are from the perspective of project managers or senior developers instead of non-contributors.

Besides user studies, there have been a lot of empirical studies on software build history and build failures. Mcintosh et al. studied the version histories of proprietary and open source software projects to estimate the effort required to repair build scripts~\cite{mcintosh2011empirical} and to correlate effort with type of bulid systems~\cite{mcintosh2015large}. Xia et al.~\cite{xia2014empirical} performed studies to summarize bugs in software build systems. Zhao et al.~\cite{zhao2014empirical} studied build failure reports in five OSS and found that build failures take much more time to fix than others.
Barrak et al.~\cite{barrak2021builds} studied how build failures are correlated with code smells in build scripts and code.
Licker and Rice~\cite{licker2019detecting} investigated the incorrect rules in build scripts by using a mutation testing approach.
Wu et al.~\cite{wu2020empirical} studied the characteristics of build failures in the context of docker environments.
These existing studies focus on build failure logs and build failure fixes in the commit history. In contrast to previous studies, our research collects and analyzes the entire process of completing multiple OSS build tasks in a local environment, taking into account the system environment factors that influence the build process.

\vspace{-.2cm}
\section{Study Design}
We conducted a study that gathered data on build issues from 31 participants involved in 12 OSS projects. This provided us with a comprehensive understanding, enabling us to identify key symptoms and examine their resolution process. We aim to answer the following research question:

\textbf{What are the common symptoms of build issues experienced by non-contributors during the build process, and to what extent can these symptoms be mitigated?}

\subsection{Participants and Tasks}
Our participants were 31 students enrolled in the same software testing course. This course is a cross-listed elective for senior-year undergraduates and graduate students, with software engineering as a prerequisite. The course, taught by two of the authors at a university, focuses on software testing approaches, test planning, test case design, and build systems with CI/CD concepts. In our study, students shared many characteristics with non-contributors who only needed to build the released OSS without modifying the source code. None of the students have the experience to set up the specific build environments required by the OSS projects in our study. Almost all of the students will work or are already working part-time or full-time as developers, so studying their behavior could further help us understand the build issues when newcomers are onboarding.

\begin{table}[bp]
  \centering
  \vspace{-.3cm}
  \caption{Selected OSS projects}
    \resizebox{0.8\linewidth}{!}{\begin{tabular}{lrrr}
    \toprule
    \multicolumn{1}{c}{PL} & \multicolumn{1}{c}{Project} & \multicolumn{1}{c}{No. Participants} & \multicolumn{1}{c}{No. Issues} \\
    \midrule
    \midrule
    C++   & \multicolumn{1}{l}{opencv/opencv} & 3     & 30 \\
          & \multicolumn{1}{l}{tensorflow/tensorflow} & 3     & 38 \\
    \midrule
    Go    & \multicolumn{1}{l}{gohugoio/hugo} & 4     & 32 \\
          & \multicolumn{1}{l}{kubernetes/kubernetes} & 2     & 18 \\
    \midrule
    Java  & \multicolumn{1}{l}{elastic/elasticsearch} & 2     & 18 \\
          & \multicolumn{1}{l}{spring-projects/spring-boot} & 3     & 46 \\
    \midrule
    JavaScript & \multicolumn{1}{l}{electron/electron} & 1     & 4 \\
          & \multicolumn{1}{l}{vuejs/vue} & 4     & 36 \\
    \midrule
    PHP   & \multicolumn{1}{l}{fzaninotto/Faker} & 2     & 9 \\
          & \multicolumn{1}{l}{guzzle/guzzle} & 1     & 15 \\
    \midrule
    Python & \multicolumn{1}{l}{pallets/flask} & 3     & 26 \\
          & \multicolumn{1}{l}{scikit-learn/scikit-learn} & 3     & 31 \\
    \midrule
    \textbf{Total} &       & \textbf{31} & \textbf{303} \\
    \end{tabular}}%
    \vspace{-0.4cm}
  \label{tab:19}%
\end{table}%

As shown in Table~\ref{tab:19}, we selected 6 popular PLs from the top 15 used on GitHub. For each PL, we chose 2 of the top 10 most popular OSS projects on GitHub, ranked by the number of stars. We made our best effort to choose PLs and OSS projects with distinct real-world application scenarios. Participants picked a project from the 12 OSS projects on a first-come, first-served basis. We limited each project to three slots to ensure that each project had at least one participant. Participants were instructed to write a report documenting at least 10 non-trivial build issues. They were encouraged to resolve the issues if they could.

Students were required to complete the study as one of their major course projects. We designed a three-stage task list to help participants become familiar with the study process:
1) Setup. Participants were instructed to use the Google Cloud Platform (GCP) online console to set up a GCP project with \$50 education credits and enabled APIs for creating virtual machines (VMs).
2) Warm-up. Participants were instructed to create a warm-up VM instance. They then performed warm-up build activities to familiarize themselves with the logging and snapshot-taking processes.
3) Build OSS projects. Each of the participants picked one OSS project for their build tasks. They utilized log scripts, snapshots, and textual issue reports to document build issues encountered during the build process of the target OSS project.

The anticipated outcomes of the study included successfully compiling the OSS projects and passing all tests using the commands specified in the build tasks. Beyond the build outcome, participants' grades would be evaluated based on their effort invested in completing the tasks, as evidenced by snapshots and issue reports. This grading approach was designed to accommodate participants who might not be able to successfully build the OSS projects. As the conductors of the study, our role was to provide clarity on the objectives of the build tasks. Our study protocol underwent assessment and received approval from our local Institutional Review Board (IRB). All students were informed that they may withdraw their data after grading, ensuring their data would not be included in our dataset for this study or any future research activities.

\vspace{-.2cm}

\subsection{Data Collection}
We designed protocols to collect data throughout the build process. As participants performed build tasks on a virtual machine (VM), we used the following methods to capture all relevant data. 

\textbf{VM state logging.}
When encountering build issues, participants used a script to log command history, environment variables, and network information. This logging helped correlate system data with reported issues. Participants also logged the final VM state upon completing tasks, with all logs saved for later analysis.

\textbf{VM snapshots.}
Snapshots captured the entire VM state at specific moments, including files, settings, and configurations. Participants manually took snapshots upon encountering build issues and upon task completion, though temporary environment variables could be lost in the process.

\textbf{Build issue report.}
Participants documented build issues, including inputs, outputs, and solutions, along with details about the build environment (e.g., language versions, tools). They summarized their process and provided feedback, offering insights into their problem-solving strategies.

\begin{figure}[bp]
\vspace{-0.4cm}
\centerline{
\includegraphics[width=\linewidth]{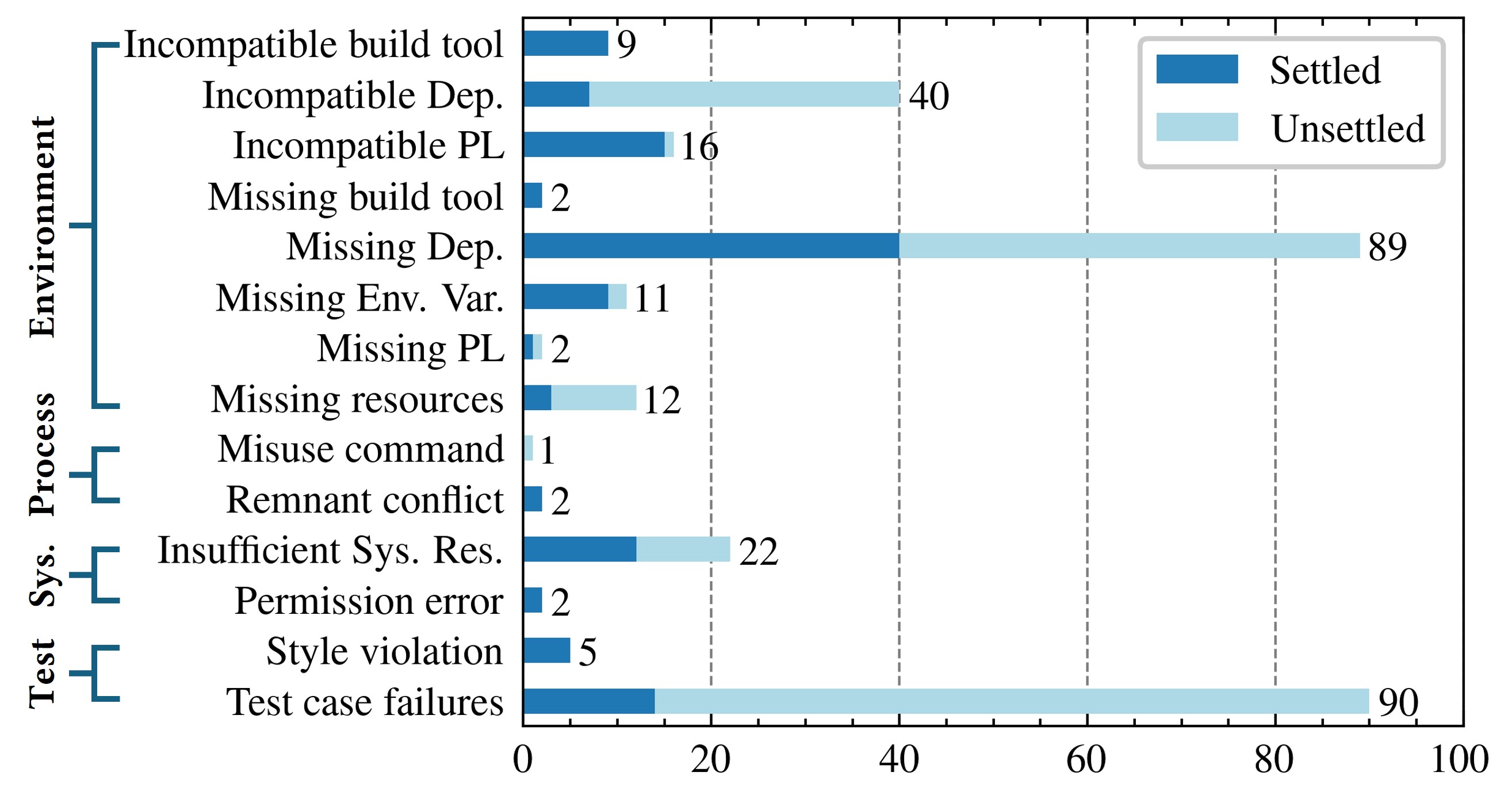}
}
\vspace{-0.4cm}
\caption{Symptoms experienced by non-contributors.}
\Description{Identified 14 distinct symptoms of build issues experienced by non-contributors in this study.}
\label{fig:symptoms19}
\end{figure}

\vspace{-.2cm}

\section{Results and Analysis}
\label{sec:results19}
We collected a total of 303 build issues from 31 build issue reports, along with 380 snapshots containing environment logs. One student changed from the ``electron/electron'' project to ``vuejs/vue'' due to low GCP credit. One student worked on the wrong project in Go. Eventually, for each OSS, there was at least one student and a maximum of four students. For each PL, there were at least three students and a maximum of six students. In the collected 303 build issues, the minimum number of issues reported was 4 for the ``electron/electron'' project, with only 1 student participant. The maximum number of issues reported was 46 for the sprint-boot project, with 3 participants. The number of reported issues was not correlated with only the number of participants but was also project-specific.

We implemented a qualitative analysis on these 303 build issues, utilizing the open coding procedure~\cite{Seaman_1999}. This procedure was executed by two of the authors. The first author coded all build issues, drawing from error messages and solutions reported within issue reports, and identified any borderline cases. Subsequently, the second author validated this coding and engaged in discussions about any borderline cases. In instances where conflicts arose, the first author leveraged corresponding snapshots and collected environment information to attempt to reproduce the build issue. Decisions were made based on the reproduced build results or the history available in the snapshot, with the two authors discussing and reaching consensus. Additionally, two authors were responsible for labeling the resolution state of the issues. If no solution was provided, the issue was labeled as \textit{Unsettled}. However, if either of the two authors believed that a solution or workaround was provided, the issue was labeled as \textit{Settled}.

Figure~\ref{fig:symptoms19} illustrates the 14 distinct symptoms of build issues identified in our study. The light-colored bars represent unsettled build issues, while the remaining dark-colored bars denote settled issues. Out of the total 303 issues analyzed, 182 (60.1\%) remained unsettled, highlighting the persistent challenges faced in addressing build issues experienced by non-contributors. To better understand the characteristics of these symptoms, we further categorized them into four broad categories: environment, process, system, and test-related issues. By grouping the 14 symptoms into these categories, we aim to provide a structured analysis of the underlying factors contributing to build issues. In the following sections, we examine each category, analyzing the symptoms and their implications for build issue resolution.

\textbf{Environment-related symptoms} include those related to incompatibility and missing components, which together account for the majority (181 out of 303) of build issues. Incompatibility-related symptoms, such as those involving build tools, dependencies (Dep.), and PL, occur when OSS projects cannot work with necessary components. Missing-related symptoms arise when build tools, dependencies, environment (Env.) variables (Var.), PLs, and external resources are not available in the current workspace. These two dependency-related symptoms have a high unsettled rate, a finding that aligns with studies of contributors-experienced build issues~\cite{lou2020understanding,Tufano_Palomba_Bavota_Di2017,Sulír_Porubän_2016,Zolfagharinia_Adams_Guéhéneuc_2019}. Missing resources, common in evolving OSS systems, are challenging for non-contributors to resolve due to the difficulty in fetching external resources.
The incompatible PL has a good settlement rate, primarily because the error messages in selected projects are easy to identify, unlike those for incompatible dependencies. Missing environment variables may mask other symptoms, as those issues require specific environment variables. We observed during the coding process that many incompatibility symptoms can be converted to other symptoms as non-contributors seek solutions.

\textbf{Process-related symptoms} refer to build issues caused by participants' mistakes during key steps of the build process. For example, ``misuse command'' symptom can occur when participants fail to escape special characters in command options, and ``remnant conflict'' symptom can arise when participants neglect to clean up the build remnants. Resolving these issues can be complex if participants are unable to correctly interpret the error messages, and the difficulty can be compounded by the complexity of the required build process.

\textbf{System-related symptoms} refer to ``Insufficient system (Sys.) resources (Res.)'' and ``Permission error''. The insufficient system resources symptom ranks among the top four symptoms, as most of the selected OSS projects are real-world projects that require a larger amount of system resources than our participants anticipated. Moreover, non-contributors often lack an effective approach to estimating the system resources required before they perform the build activities. This could explain the high unsettled rate of 10 out of 22. This symptom is also common since VMs created by our participants typically do not have many redundant system resources. For instance, participants experienced system crashes due to disk space exhaustion, followed by failures to reconnect to VMs. They did not realize the issue until the system resources were exhausted. Another symptom is the ``permission error'', which is caused when underlying software requires higher permissions than it currently has. In our study, we found that \textit{Docker} commands often need superuser permissions. This is a common issue among OSS projects that rely on \textit{Docker}~\cite{docker}.

\textbf{Test-related symptoms} refer to ``Style violations'' and ``Test case failures''. Style violation test failures typically occur when a project does not adhere to a certain standard. Participants can often find a workaround to bypass this issue. However, it is important to note that the root cause of this symptom may be due to compatibility issues.
The most common symptom in our study is the ``Test case failures'' symptom, which also has a high unsettled rate. This aligns with existing studies~\cite{Rausch_Hummer_Leitner_Schulte_2017,Vassallo_Schermann_Zampetti_Romano_Leitner_Zaidman_Di}. Notably, some test failure messages are difficult to interpret, making it hard to identify the root cause. The symptom could be the result of other underlying issues.

\begin{figure}[tbp]
\centerline{
\includegraphics[width=0.65\linewidth]{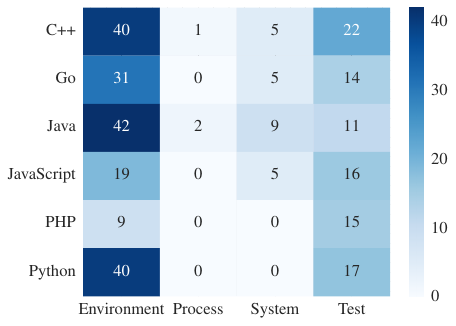}
}
\vspace{-0.4cm}

\caption{Distribution of categories on PLs.}
\Description{Distribution of categories across different PLs.}
\vspace{-0.1cm}

\label{fig:PL_heatmap}
\vspace{-0.5cm}
\end{figure}

The distribution of these categories among the participants is as follows: ``Environment'' was experienced by 29 participants (93.6\%), ``Process'' by 2 participants (6.5\%), ``System'' by 10 participants (32.3\%), and ``Test'' by 24 participants (77.4\%). Figure~\ref{fig:PL_heatmap} illustrates the distribution of categories across different PLs. Environment and test related symptoms are present in all PLs. In contrast, system-related symptoms are project-specific, which can be attributed to the fact that lightweight projects do not consume significant system resources. Upon further investigation of process-related symptoms, we observed that a participant encountered both ``Misuse command'' and ``Remnant conflict'' issues in a Spring Boot project. Despite its infrequent occurrence among the issues observed, the ``Remnant conflict'' issue was identified in two projects, one written in Java and the other in C++.

Compared to the study by Lou et al.~\cite{lou2020understanding}, our study on the symptoms of build issues shows that non-contributors still frequently experience the same symptoms as contributors. However, these are mostly limited to issues caused by a lack of installation, version incompatibility errors, and environment variable issues. We also observed additional symptoms not covered by previous studies that are related to participants' actions, such as ``Command misuse'' and ``remnant conflicts''. As the 39.9\% settlement rate shows, participants found it difficult to mitigate most of these frequent symptoms as well as identify the root causes of build issues. The strategies for resolving these build issues tend to be more straightforward due to the non-contributors' lack of specific project knowledge and experience. This limitation can make them more susceptible to encountering certain build issues.

\vspace{-0.2cm}
\section{Discussion and Future Work}
\label{sec:discuss}

Although containerization holds promise for addressing certain build issues, the lack of accessible container environments in non-contributors' local systems presented a challenge for many OSS projects, including most of those in our study.
A threat to the validity of the findings is the subjectivity involved in determining whether a build issue is trivial. This judgment depends on several factors, including the individual's level of knowledge and local systems. Our future research will involve more participants and focus on a specific scope to explore the correlation between various factors and our findings. 
Additionally, we observed inconsistencies in reporting, such as misinterpreted test results and misaligned snapshots. Therefore, incorporating more monitoring measures into the process could significantly mitigate these potential threats.

\vspace{-0.2cm}

\section{Conclusion}
\label{sec:conclude}
This paper investigates 303 build issues experienced by 31 non-contributors. We found that non-contributors often struggle to resolve issues due to limited symptoms. Our study provides valuable insights into build issue resolution from the perspective of non-contributors, highlighting the importance of understanding their behavior and challenges. This work lays the foundation for further research in this area, with the ultimate goal of improving the experience of non-contributors in dealing build issues.\footnote{This work is supported in part by NSF projects 1736209 and 1846467.}
\balance
\bibliographystyle{ACM-Reference-Format}
\bibliography{refs}


\begin{thebibliography}{25}


\ifx \showCODEN    \undefined \def \showCODEN     #1{\unskip}     \fi
\ifx \showDOI      \undefined \def \showDOI       #1{#1}\fi
\ifx \showISBNx    \undefined \def \showISBNx     #1{\unskip}     \fi
\ifx \showISBNxiii \undefined \def \showISBNxiii  #1{\unskip}     \fi
\ifx \showISSN     \undefined \def \showISSN      #1{\unskip}     \fi
\ifx \showLCCN     \undefined \def \showLCCN      #1{\unskip}     \fi
\ifx \shownote     \undefined \def \shownote      #1{#1}          \fi
\ifx \showarticletitle \undefined \def \showarticletitle #1{#1}   \fi
\ifx \showURL      \undefined \def \showURL       {\relax}        \fi
\providecommand\bibfield[2]{#2}
\providecommand\bibinfo[2]{#2}
\providecommand\natexlab[1]{#1}
\providecommand\showeprint[2][]{arXiv:#2}

\bibitem[Bit(2017)]%
        {Bitcoin}
 \bibinfo{year}{2017}\natexlab{}.
\newblock \bibinfo{title}{Difference between downloading bitcoire core from bitcoin.org and compiling from Github}.
\newblock \bibinfo{howpublished}{\url{https://bitcoin.stackexchange.com/questions/59875}}.
\newblock
\newblock
\shownote{Accessed: 2024-03-21}.


\bibitem[doc(2018)]%
        {docker}
 \bibinfo{year}{2018}\natexlab{}.
\newblock \bibinfo{title}{How to fix docker: Got permission denied issue}.
\newblock \bibinfo{howpublished}{\url{stackoverflow.com/questions/48957195}}.
\newblock
\newblock
\shownote{Accessed: 2024}.


\bibitem[Barrak et~al\mbox{.}(2021)]%
        {barrak2021builds}
\bibfield{author}{\bibinfo{person}{Amine Barrak}, \bibinfo{person}{Ellis~E Eghan}, \bibinfo{person}{Bram Adams}, {and} \bibinfo{person}{Foutse Khomh}.} \bibinfo{year}{2021}\natexlab{}.
\newblock \showarticletitle{Why do builds fail?—A conceptual replication study}.
\newblock \bibinfo{journal}{\emph{Journal of Systems and Software}}  \bibinfo{volume}{177} (\bibinfo{year}{2021}), \bibinfo{pages}{110939}.
\newblock


\bibitem[Downs et~al\mbox{.}(2012)]%
        {downs2012ambient}
\bibfield{author}{\bibinfo{person}{John Downs}, \bibinfo{person}{Beryl Plimmer}, {and} \bibinfo{person}{John~G Hosking}.} \bibinfo{year}{2012}\natexlab{}.
\newblock \showarticletitle{Ambient awareness of build status in collocated software teams}. In \bibinfo{booktitle}{\emph{2012 34th International Conference on Software Engineering (ICSE)}}. IEEE, \bibinfo{pages}{507--517}.
\newblock


\bibitem[Hilton et~al\mbox{.}(2016)]%
        {hilton2016continuous}
\bibfield{author}{\bibinfo{person}{Michael Hilton}, \bibinfo{person}{Nicholas Nelson}, \bibinfo{person}{Danny Dig}, \bibinfo{person}{Timothy Tunnell}, \bibinfo{person}{Darko Marinov}, {et~al\mbox{.}}} \bibinfo{year}{2016}\natexlab{}.
\newblock \showarticletitle{Continuous integration (CI) needs and wishes for developers of proprietary code}.
\newblock  (\bibinfo{year}{2016}).
\newblock


\bibitem[Kerzazi et~al\mbox{.}(2014)]%
        {kerzazi2014automated}
\bibfield{author}{\bibinfo{person}{Noureddine Kerzazi}, \bibinfo{person}{Foutse Khomh}, {and} \bibinfo{person}{Bram Adams}.} \bibinfo{year}{2014}\natexlab{}.
\newblock \showarticletitle{Why do automated builds break? an empirical study}. In \bibinfo{booktitle}{\emph{2014 IEEE International Conference on Software Maintenance and Evolution}}. IEEE, \bibinfo{pages}{41--50}.
\newblock


\bibitem[Kwan et~al\mbox{.}(2011)]%
        {kwan2011does}
\bibfield{author}{\bibinfo{person}{Irwin Kwan}, \bibinfo{person}{Adrian Schroter}, {and} \bibinfo{person}{Daniela Damian}.} \bibinfo{year}{2011}\natexlab{}.
\newblock \showarticletitle{Does socio-technical congruence have an effect on software build success? a study of coordination in a software project}.
\newblock \bibinfo{journal}{\emph{IEEE Transactions on Software Engineering}} \bibinfo{volume}{37}, \bibinfo{number}{3} (\bibinfo{year}{2011}), \bibinfo{pages}{307--324}.
\newblock


\bibitem[Licker and Rice(2019)]%
        {licker2019detecting}
\bibfield{author}{\bibinfo{person}{N{\'a}ndor Licker} {and} \bibinfo{person}{Andrew Rice}.} \bibinfo{year}{2019}\natexlab{}.
\newblock \showarticletitle{Detecting incorrect build rules}. In \bibinfo{booktitle}{\emph{2019 IEEE/ACM 41st International Conference on Software Engineering (ICSE)}}. IEEE, \bibinfo{pages}{1234--1244}.
\newblock


\bibitem[Lou et~al\mbox{.}(2020)]%
        {lou2020understanding}
\bibfield{author}{\bibinfo{person}{Yiling Lou}, \bibinfo{person}{Zhenpeng Chen}, \bibinfo{person}{Yanbin Cao}, \bibinfo{person}{Dan Hao}, {and} \bibinfo{person}{Lu Zhang}.} \bibinfo{year}{2020}\natexlab{}.
\newblock \showarticletitle{Understanding build issue resolution in practice: symptoms and fix patterns}. In \bibinfo{booktitle}{\emph{Proceedings of the 28th ACM Joint Meeting on European Software Engineering Conference and Symposium on the Foundations of Software Engineering}}. \bibinfo{pages}{617--628}.
\newblock


\bibitem[McIntosh et~al\mbox{.}(2011)]%
        {mcintosh2011empirical}
\bibfield{author}{\bibinfo{person}{Shane McIntosh}, \bibinfo{person}{Bram Adams}, \bibinfo{person}{Thanh~HD Nguyen}, \bibinfo{person}{Yasutaka Kamei}, {and} \bibinfo{person}{Ahmed~E Hassan}.} \bibinfo{year}{2011}\natexlab{}.
\newblock \showarticletitle{An empirical study of build maintenance effort}. In \bibinfo{booktitle}{\emph{Proceedings of the 33rd international conference on software engineering}}. \bibinfo{pages}{141--150}.
\newblock


\bibitem[McIntosh et~al\mbox{.}(2015)]%
        {mcintosh2015large}
\bibfield{author}{\bibinfo{person}{Shane McIntosh}, \bibinfo{person}{Meiyappan Nagappan}, \bibinfo{person}{Bram Adams}, \bibinfo{person}{Audris Mockus}, {and} \bibinfo{person}{Ahmed~E Hassan}.} \bibinfo{year}{2015}\natexlab{}.
\newblock \showarticletitle{A large-scale empirical study of the relationship between build technology and build maintenance}.
\newblock \bibinfo{journal}{\emph{Empirical Software Engineering}}  \bibinfo{volume}{20} (\bibinfo{year}{2015}), \bibinfo{pages}{1587--1633}.
\newblock


\bibitem[Overflow(2024)]%
        {SO}
\bibfield{author}{\bibinfo{person}{Stack Overflow}.} \bibinfo{year}{2024}\natexlab{}.
\newblock \bibinfo{title}{Stack Overflow: a question-and-answer website for computer programmers}.
\newblock
\newblock
\urldef\tempurl%
\url{https://stackoverflow.com/}
\showURL{%
\tempurl}


\bibitem[Phillips et~al\mbox{.}(2014)]%
        {phillips2014understanding}
\bibfield{author}{\bibinfo{person}{Shaun Phillips}, \bibinfo{person}{Thomas Zimmermann}, {and} \bibinfo{person}{Christian Bird}.} \bibinfo{year}{2014}\natexlab{}.
\newblock \showarticletitle{Understanding and improving software build teams}. In \bibinfo{booktitle}{\emph{Proceedings of the 36th international conference on software engineering}}. \bibinfo{pages}{735--744}.
\newblock


\bibitem[Rausch et~al\mbox{.}(2017)]%
        {Rausch_Hummer_Leitner_Schulte_2017}
\bibfield{author}{\bibinfo{person}{Thomas Rausch}, \bibinfo{person}{Waldemar Hummer}, \bibinfo{person}{Philipp Leitner}, {and} \bibinfo{person}{Stefan Schulte}.} \bibinfo{year}{2017}\natexlab{}.
\newblock \showarticletitle{An empirical analysis of build failures in the continuous integration workflows of java-based open-source software}. In \bibinfo{booktitle}{\emph{2017 IEEE/ACM 14th International Conference on Mining Software Repositories (MSR)}}. \bibinfo{publisher}{IEEE}, \bibinfo{pages}{345–355}.
\newblock
\urldef\tempurl%
\url{https://ieeexplore.ieee.org/abstract/document/7962384/}
\showURL{%
\tempurl}


\bibitem[Seaman(1999)]%
        {Seaman_1999}
\bibfield{author}{\bibinfo{person}{C.B. Seaman}.} \bibinfo{year}{1999}\natexlab{}.
\newblock \showarticletitle{Qualitative methods in empirical studies of software engineering}.
\newblock \bibinfo{journal}{\emph{IEEE Transactions on Software Engineering}} \bibinfo{volume}{25}, \bibinfo{number}{4} (\bibinfo{date}{July} \bibinfo{year}{1999}), \bibinfo{pages}{557–572}.
\newblock
\showISSN{1939-3520}
\urldef\tempurl%
\url{https://doi.org/10.1109/32.799955}
\showDOI{\tempurl}


\bibitem[Shridhar et~al\mbox{.}(2014)]%
        {shridhar2014qualitative}
\bibfield{author}{\bibinfo{person}{Mini Shridhar}, \bibinfo{person}{Bram Adams}, {and} \bibinfo{person}{Foutse Khomh}.} \bibinfo{year}{2014}\natexlab{}.
\newblock \showarticletitle{A qualitative analysis of software build system changes and build ownership styles}. In \bibinfo{booktitle}{\emph{Proceedings of the 8th ACM/IEEE international symposium on empirical software engineering and measurement}}. \bibinfo{pages}{1--10}.
\newblock


\bibitem[Sulír and Porubän(2016)]%
        {Sulír_Porubän_2016}
\bibfield{author}{\bibinfo{person}{Matúš Sulír} {and} \bibinfo{person}{Jaroslav Porubän}.} \bibinfo{year}{2016}\natexlab{}.
\newblock \showarticletitle{A quantitative study of Java software buildability}. In \bibinfo{booktitle}{\emph{Proceedings of the 7th International Workshop on Evaluation and Usability of Programming Languages and Tools}}. \bibinfo{publisher}{ACM}, \bibinfo{address}{Amsterdam Netherlands}, \bibinfo{pages}{17–25}.
\newblock
\showISBNx{978-1-4503-4638-2}
\urldef\tempurl%
\url{https://doi.org/10.1145/3001878.3001882}
\showDOI{\tempurl}


\bibitem[Tufano et~al\mbox{.}(2017)]%
        {Tufano_Palomba_Bavota_Di2017}
\bibfield{author}{\bibinfo{person}{Michele Tufano}, \bibinfo{person}{Fabio Palomba}, \bibinfo{person}{Gabriele Bavota}, \bibinfo{person}{Massimiliano Di Penta}, \bibinfo{person}{Rocco Oliveto}, \bibinfo{person}{Andrea De Lucia}, {and} \bibinfo{person}{Denys Poshyvanyk}.} \bibinfo{year}{2017}\natexlab{}.
\newblock \showarticletitle{There and back again: Can you compile that snapshot?}
\newblock \bibinfo{journal}{\emph{Journal of Software: Evolution and Process}} \bibinfo{volume}{29}, \bibinfo{number}{4} (\bibinfo{date}{April} \bibinfo{year}{2017}), \bibinfo{pages}{e1838}.
\newblock
\showISSN{2047-7473, 2047-7481}
\urldef\tempurl%
\url{https://doi.org/10.1002/smr.1838}
\showDOI{\tempurl}


\bibitem[Vassallo et~al\mbox{.}(2020)]%
        {vassallo2020every}
\bibfield{author}{\bibinfo{person}{Carmine Vassallo}, \bibinfo{person}{Sebastian Proksch}, \bibinfo{person}{Timothy Zemp}, {and} \bibinfo{person}{Harald~C Gall}.} \bibinfo{year}{2020}\natexlab{}.
\newblock \showarticletitle{Every build you break: developer-oriented assistance for build failure resolution}.
\newblock \bibinfo{journal}{\emph{Empirical Software Engineering}}  \bibinfo{volume}{25} (\bibinfo{year}{2020}), \bibinfo{pages}{2218--2257}.
\newblock


\bibitem[Vassallo et~al\mbox{.}(2017)]%
        {Vassallo_Schermann_Zampetti_Romano_Leitner_Zaidman_Di}
\bibfield{author}{\bibinfo{person}{Carmine Vassallo}, \bibinfo{person}{Gerald Schermann}, \bibinfo{person}{Fiorella Zampetti}, \bibinfo{person}{Daniele Romano}, \bibinfo{person}{Philipp Leitner}, \bibinfo{person}{Andy Zaidman}, \bibinfo{person}{Massimiliano Di~Penta}, {and} \bibinfo{person}{Sebastiano Panichella}.} \bibinfo{year}{2017}\natexlab{}.
\newblock \showarticletitle{A tale of CI build failures: An open source and a financial organization perspective}. In \bibinfo{booktitle}{\emph{2017 IEEE international conference on software maintenance and evolution (ICSME)}}. \bibinfo{publisher}{IEEE}, \bibinfo{pages}{183–193}.
\newblock
\urldef\tempurl%
\url{https://ieeexplore.ieee.org/abstract/document/8094420/}
\showURL{%
\tempurl}


\bibitem[Wu et~al\mbox{.}(2020)]%
        {wu2020empirical}
\bibfield{author}{\bibinfo{person}{Yiwen Wu}, \bibinfo{person}{Yang Zhang}, \bibinfo{person}{Tao Wang}, {and} \bibinfo{person}{Huaimin Wang}.} \bibinfo{year}{2020}\natexlab{}.
\newblock \showarticletitle{An empirical study of build failures in the docker context}. In \bibinfo{booktitle}{\emph{Proceedings of the 17th international conference on mining software repositories}}. \bibinfo{pages}{76--80}.
\newblock


\bibitem[Xia et~al\mbox{.}(2014)]%
        {xia2014empirical}
\bibfield{author}{\bibinfo{person}{Xin Xia}, \bibinfo{person}{Xiaozhen Zhou}, \bibinfo{person}{David Lo}, \bibinfo{person}{Xiaoqiong Zhao}, {and} \bibinfo{person}{Ye Wang}.} \bibinfo{year}{2014}\natexlab{}.
\newblock \showarticletitle{An empirical study of bugs in software build system}.
\newblock \bibinfo{journal}{\emph{IEICE TRANSACTIONS on Information and Systems}} \bibinfo{volume}{97}, \bibinfo{number}{7} (\bibinfo{year}{2014}), \bibinfo{pages}{1769--1780}.
\newblock


\bibitem[Zhao et~al\mbox{.}(2014)]%
        {zhao2014empirical}
\bibfield{author}{\bibinfo{person}{Xiaoqiong Zhao}, \bibinfo{person}{Xin Xia}, \bibinfo{person}{Pavneet~Singh Kochhar}, \bibinfo{person}{David Lo}, {and} \bibinfo{person}{Shanping Li}.} \bibinfo{year}{2014}\natexlab{}.
\newblock \showarticletitle{An empirical study of bugs in build process}. In \bibinfo{booktitle}{\emph{Proceedings of the 29th Annual ACM Symposium on Applied Computing}}. \bibinfo{pages}{1187--1189}.
\newblock


\bibitem[Zolfagharinia et~al\mbox{.}(2017)]%
        {zolfagharinia2017not}
\bibfield{author}{\bibinfo{person}{Mahdis Zolfagharinia}, \bibinfo{person}{Bram Adams}, {and} \bibinfo{person}{Yann-Ga{\"e}l Gu{\'e}h{\'e}nuc}.} \bibinfo{year}{2017}\natexlab{}.
\newblock \showarticletitle{Do not trust build results at face value-an empirical study of 30 million cpan builds}. In \bibinfo{booktitle}{\emph{2017 IEEE/ACM 14th International Conference on Mining Software Repositories (MSR)}}. IEEE, \bibinfo{pages}{312--322}.
\newblock


\bibitem[Zolfagharinia et~al\mbox{.}(2019)]%
        {Zolfagharinia_Adams_Guéhéneuc_2019}
\bibfield{author}{\bibinfo{person}{Mahdis Zolfagharinia}, \bibinfo{person}{Bram Adams}, {and} \bibinfo{person}{Yann-Gaël Guéhéneuc}.} \bibinfo{year}{2019}\natexlab{}.
\newblock \showarticletitle{A study of build inflation in 30 million CPAN builds on 13 Perl versions and 10 operating systems}.
\newblock \bibinfo{journal}{\emph{Empirical Software Engineering}} \bibinfo{volume}{24}, \bibinfo{number}{6} (\bibinfo{year}{2019}), \bibinfo{pages}{3933–3971}.
\newblock


\end{thebibliography}

\end{document}